\documentclass[10pt,letterpaper]{article}

\usepackage{authblk}
\usepackage{amsmath,amsfonts,amsthm,amssymb,algorithm,algpseudocode}
\usepackage{bm,booktabs}
\usepackage{color}
\usepackage{dashrule}
\usepackage{enumitem}
\usepackage{float,fancyhdr}
\usepackage{graphicx,graphics}
\usepackage{indentfirst}
\usepackage{mathrsfs,multirow,multicol}
\usepackage{nicefrac}
\usepackage{setspace,subfigure}
\usepackage{tabularx,tabu}
\usepackage[super]{natbib}
\usepackage[letterpaper, margin=1in]{geometry}
\usepackage[colorlinks=true]{hyperref}

\usepackage[latin1]{inputenc}
\usepackage[english]{babel}

\title{wTO: an R package for computing weighted topological overlap and consensus networks with an integrated visualization tool}
\author[1,2,*]{Deisy Morselli Gysi}
\author[3]  {Andre Voigt}
\author[4]  {Tiago Miranda Fragoso} 
\author[3,5]  {Eivind Almaas} 
\author[6]{Katja Nowick} 

\affil[1]{Department of Computer Science, Interdisciplinary Center of Bioinformatics, University of Leipzig, Leipzig, D-04109, Leipzig.}
\affil[2]{Swarm Intelligence and Complex Systems Group, Faculty of Mathematics and Computer Science, University of Leipzig,  Leipzig, D-04109, Leipzig.}
\affil[3]{Department of Biotechnology,  NTNU - Norwegian University of Science and Technology, Trondheim, Norway.}
\affil[4]{Funda\c{c}\~ao Cesgranrio, Rio de Janeiro, 20261-903, Brazil.}
\affil[5]{K.G. Jebsen Center for Genetic Epidemiology, NTNU - Norwegian University of Science and Technology, Trondheim, Norway.}
\affil[6]{Human Biology Group, Institute for Biology, Department of Biology, Chemistry, Pharmacy, Free University of Berlin, Konigin-Luise-Str. 1-3, D-14195 Berlin, Germany.}
\affil[*]{To whom correspondence should be addressed. deisy@bioinf.uni-leipzig.de}

\begin{document}
\maketitle

\begin{abstract}
\noindent \textbf{Background} { Network analyses, such as of gene co-expression networks, metabolic networks and ecological networks have become a central approach for the systems-level study of biological data}. Several software packages exist for generating and analyzing such networks, either from correlation scores or the absolute value of a transformed score called weighted topological overlap ($wTO$). However, since gene regulatory processes can up- or down-regulate genes, it is of great interest to explicitly consider both positive and negative correlations when constructing a gene co-expression network. \textbf{Results} Here, we present an \texttt{R} package for calculating the weighted topological overlap ($wTO$), that, in contrast to existing packages,  explicitly addresses  the sign of the $wTO$ values, and is thus especially valuable for the analysis of gene regulatory networks.
 The package includes the calculation of p-values (raw and adjusted) for each pairwise gene score. Our package also allows the calculation of networks from time series (without replicates). Since networks from independent datasets (biological repeats or related studies) are not the same due to technical and biological noise in the data, we additionally, incorporated a novel method for calculating a consensus network ($CN$) from two or more networks into our \texttt{R} package. To graphically inspect the resulting networks, the \texttt{R} package contains a visualization tool, which allows for the direct network manipulation and access of node and link information. When testing the package on a standard laptop computer, we can conduct all calculations for systems of more than $20,000$ genes in under two hours. {We compare our new wTO package to state of art packages and demonstrate the application of the wTO and CN functions using 3 independently derived datasets from healthy human pre-frontal cortex samples. To showcase an example for the time series application we utilized a metagenomics data set.}
\textbf{Conclusion} In this work, we developed a software package that allows the computation of $wTO$ networks, $CN$s and a visualization tool in the \texttt{R} statistical environment. It is publicly available on CRAN repositories under the GPL$-$2 Open Source License (\url{https://cran.r-project.org/web/packages/wTO/}).
\end{abstract}
\textit{Network},
\textit{Expression},
\textit{R package},
\textit{Software},
\textit{Consensus Network},
\textit{wTO}.

\section*{Background}
Recent applications of complex network analysis methods have provided important new knowledge of the functioning and interactions of genes at the systems level~\cite{barabasi2004network, bansal2007infer, furlong2013human,  dempsey2013mining}. Within the area of biological network analyses, co-expression networks have received much attention~\cite{yang2014gene,taylor2009dynamic}. For the co-expression networks, a pair of nodes are typically connected by a link if the genes they represent show a significantly correlated expression pattern. In the network, this link may be represented as a binary relationship, where $1 = $ ``presence'' and  $0 = $ ``absence'' of the link,  or alternatively, the link may have a numeric value (often called weight). The magnitude of the weight is typically interpreted as representing the strength of a gene-pair relationship, and the sign as indicative of the type of associated gene interaction: positive if the genes are co-regulated, negative if they are oppositely controlled~\cite{van2017gene}.

In many implementations of network analyses, we may primarily be interested in an {\em a priori} defined subset of genes with a specific set of properties. Examples include transcription factors (TFs), genes with known orthologs in a set of organisms of interest, or disease {associated} genes ~\cite{babu2004structure, mason2009signed}.
For these situations, oftentimes the choice is made to only take into account direct interactions between the gene-subset of interest, instead of including the full set of correlations. A major drawback with such an approach, is that relevant information contained in  interaction patterns among  excluded genes that would affect network topology and link strength values, is not incorporated in the network. The loss of such information is not only undesirable, but may also  lead to biased results. 

{When analyzing networks in which the links have non-binary weights, the method of 
weighted topological ($wTO$) network analysis~\cite{ravasz2002hierarchical} has been found very useful. In a $wTO$-analysis, a new link-weight 
for a pair of connected nodes is determined through an averaging process that accounts for {\it all} common network neighbors~\cite{ravasz2002hierarchical}.} Thus, $wTO$ is a method that {\em implicitly} includes correlations among nodes that are going to be exempt from further analysis. The $wTO$ method~\cite{ravasz2002hierarchical,zhang2005general, carlson2006gene} can be used to determine the overlap among classes of transcripts, for example TFs  and non-coding RNAs (ncRNAs). The resulting $wTO$ network provides a more robust representation of the connections and interactions among the node-set of interest than a simple correlation network analysis focused only on the node-set of interest~\cite{nowick2009differences}. 

{The packages \texttt{WGCNA}~\cite{langfelder2008wgcna, fastR} and \texttt{ARACNe}~\cite{margolin2006reverse, margolin2006ARACNe} are widely used for weighted gene co-expression network analysis studies.} {The former} provides functions for the calculation of the adjacency matrix for all pairs of genes as the n-th power of absolute correlations, resulting in an unsigned network. Network modules can be defined with this package by unsupervised clustering. {The latter uses the mutual information (MI) of the expression in order to build the networks}. {These methods have received much attention in the literature~\cite{Allen2012,van2017gene}}. 

Previously, Nowick and collaborators~\cite{nowick2009differences} developed a mathematical method to calculate the $wTO$ for a set of nodes that explicitly takes into account both positive and negative correlations. This version of the wTO-measure is especially valuable {for investigating networks, in which it matters whether an interaction is activating or inhibiting/repressing. For instance, in gene regulatory networks the effect of a transcription factor or a ncRNA on its target genes can be activating or repressing. In metabolic networks, the increase of a substance can lead to an increase or decrease of another substance. Or in ecological networks, species interactions can be positive or negative, for instance in symbiotic or predator-prey relationships. In such cases, a distinction between positive and negative correlations for the calculation of the wTO is necessary and using the absolute correlations would falsify the biological insights.} This $wTO$-calculation methodology is implemented in the \texttt{R} package presented here. In order to avoid confusion, we will refer to the method for calculating a pair-wise link score as $wTO$ and to the package as \texttt{wTO}.

When analyzing similar datasets, e.g. from a repeated experiment or independent studies on a similar subject, the resulting networks are usually different~\cite{berto2016consensus}. These differences may arise from several sources: (A) technical differences, such as the platform on which the expression data was measured, the facility where data was collected and prepared, or how data was processed. (B) Another cause may be biological differences from confounding factors, such as sex, age, and geographic origin of the individuals measured. It is thus desirable to obtain an integrated network that  considers all independently derived networks as biological replicates and systematically identifies their commonalities. We {developed} a novel method to compute the network that captures all this information; we call this the consensus network ($CN$).

\begin{figure}[ht]
	\centering
	\includegraphics[width=\textwidth]{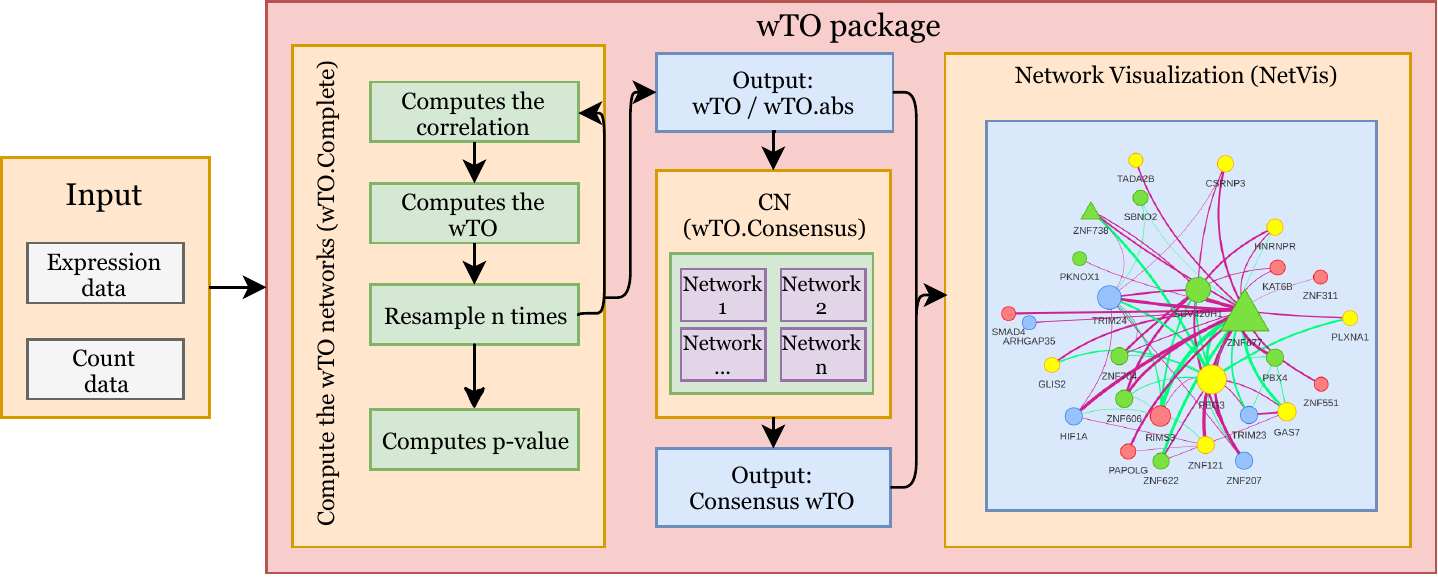}
	\caption{\textbf{The \texttt{wTO} package workflow}   Gray 
		boxes refer to inputs, red boxes refer to content of the \texttt{wTO} package, yellow boxes are functions included in the package, blue boxes are outputs of those functions, and green boxes refer to methods internal to the package. Our package can deal with multiple kinds of data, for example RNA-seq counts or normalized values, microarray expression data, abundance data coming from metagenomic studies, and many more. {All input data should be pre-processed with the quality control and normalization methods recommended for each respective type of data.} The function \texttt{wTO.Complete} calculates the $wTO$ values, as many times as desired. As output, the user will obtain an object containing the signed and absolute $wTO$ values for each pair of nodes,  $p$-values and  $p_{adj}$~values   for multiple testing. This output can be used for the construction of a $CN$ from independent networks using the function \texttt{wTO.Consensus}. Outputs from the $wTO$ and $CN$ networks can be used as an input for \texttt{NetVis}, which is an integrated tool for plotting networks. As an interactive tool it also allows the user to modify the network.}
	\label{figure:workflow}
\end{figure}

Here, we present \texttt{wTO}, an \texttt{R} package that is capable of computing both signed and unsigned $wTO$ networks as well as the $CN$, {thus providing} methods for assigning $p$-values to each link.  {The package also} comes with an integrated tool to visualize the resulting networks {and allows for nine different methods for network clustering to aid in module identification}. The workflow of the package is shown in Fig.~\ref{figure:workflow}.

{We compare our method to other state of art methods. To exemplify the usage of our package, we show here results from the calculation of wTO and CN networks from three independent genome-wide expression studies of healthy human pre-frontal cortex samples and an analysis of a time-series dataset from a metagenomics study.}

\section*{Implementation}

\subsection*{Input data}
{
Our package can handle a wide range of input data. Data can be discrete or continuous values. We recommend performing all commonly used steps for quality control and normalization before passing on the data to our package. For RNA-Seq data, our package can handle normalized quantification, for example RPKM (Reads Per Kilobase Million), FPKM (Fragments Per Kilobase Million), and TPM (Transcripts Per Kilobase Million). For microarray data, rma or mas5 values can be used. If our package is used with metagenomics data, for instance for analyzing co-occurrence networks, we recommend the abundance data to be normalized per day/ sample.}

\subsection*{Weighted Topological Overlap calculation}

For a system of N nodes ({e.g. }genes or species), we define the adjacency matrix $A = [a_{i,j}]$ based on correlations between a pair of nodes $i$ and $j$ as
\begin{align}
a_{i,j} = \begin{cases}
\rho_{i,j} & i \neq j\\ 
 0 & i = j .
\end{cases}
\end{align}
with $\rho_{i,j}$ being a correlation measure.
Assuming that nodes $i$ and $j$ {represent a sub-set of factors (e.g genes) of particular interest} selected from the $N$ nodes, 
we calculate the {weighted topological overlap ($wTO$ ~\cite{nowick2009differences}{, $\omega_{i,j}$})} between node $i$ and node $j$ as
%$\text{wTO} = [\omega_{i,j}]$, where
\begin{align}
\omega_{i,j} = \frac{\sum_{u=1}^N a_{i,u}a_{u,j} + a_{i,j}}{\min (k_i, k_j) + 1 - |a_{i,j}|},
\label{eq:wto}
\end{align}
where 
\begin{align}
k_i = \sum_{j=1}^N |a_{i,j}|.
\end{align}
Note that, 
this expression explicitly includes both positive and negative correlations, and thus allows for {$\omega_{i,j}$} to take both positive and negative values. 
Other software packages calculating the {$\omega_{i,j}$} have implemented definitions of the $wTO$ method that do not allow for negative values~\cite{langfelder2008wgcna}, making this  version  valuable for gene regulatory network analysis. {The wTO package also calculates the unsigned network, and for that, it takes as an input the absolute values of the correlation.}

Since Eq.~(\ref{eq:wto}) explicitly allows $a_{i,j} \leqslant 0$, we need to be aware of the limits of this expression. Consider three nodes $i$, $j$ and $u$, and assume that $a_{ij} \leqslant 0$.  All the terms in the {numerator} of Eq.~(\ref{eq:wto}) will be negative if $a_{iu}a_{uj} \leqslant 0$ for all nodes $u$. However, if $a_{iu}a_{uj} > 0$, then at least some contributions to the sum will cancel out. 
{The same rationale applies} for the case of $a_{ij} \geq 0$. 

To systematically assess the potential effect of term cancellation in Eq.~(\ref{eq:wto}), we calculate the absolute weighted topological overlap, {$|\omega|$} which uses the absolute value of the correlations ($a_{i,j}=|a_{i,j}|$) as input for Eq.~(\ref{eq:wto}). 
In this case, the sign of the correlation is excluded from the analysis and only the magnitude of the link-strength is taken into account. 
Consequently, by generating a scatter plot of the signed and unsigned {weights}, it is possible to assess at which {$\omega_{i,j}$}-values term cancellations start affecting the results. Thus, for $wTO$ values of interest, the closer the plot of {$\omega$} vs. {$|\omega|$} is to $y = |x|$, the better.

However, by just computing the $wTO$ network we do not avoid all spurious correlations. A way to detect them is to compute a probability of each one of the link scores being zero {using the hypothesis test
\begin{align}
\begin{cases}
H_0: \omega_{ij} = 0 \\
H_a: \omega_{ij} \neq 0
\end{cases},
\end{align}}
\noindent
{of the null hypothesis ($H_0$) of no association against the two-sided alternative ($H_a$) of non-zero association. This  can be computed by using bootstrap \cite{efron1994introduction}  or permutation resampling methods \cite{nowick2009differences}. In the former, one resamples individuals, thus approximating the weights' empirical distribution and calculating the probability that an observed weight is sufficiently distant from zero. 
In the latter, one operates under the null hypothesis of no dependence among genes and permutes the gene labels, obtaining the weights' distribution under the null hypothesis, which is rejected if the observed weight is sufficiently extreme.} {We define $\delta$ as the maximal distance between the $\omega_{i,j}$ calculated with each bootstrap and the $\omega_{i,j}$ of the real dataset. This means that, the smaller $\delta$ is, the stronger is our confidence in a particular $\omega_{i,j}$. By default, $\delta$ is set to 0.2.}

{One advantage of the wTO package is its application to analyze and make networks out of time-series data. Therefore, we are interested in the implementation of blocked bootstrap resampling~\cite{efron1994introduction} that can be used for temporal data without sample replicates for each time point. This type of resampling is necessary once there are two correlation components in those samples: The correlation inside the factors of each sample and the correlation across the time of different samples. For this situation, the use of a lag is required. 
Lags are particularly helpful in time-series analyses as autocorrelations are often present: a tendency of consecutive values to be correlated. An important benefit of the presence of autocorrelations is that we may be able to identify patterns inside a time-series, such as seasonality (patterns that repeat themselves at a periodic frequency). Therefore, the lag can be chosen using a partial correlation of the time per sample. This is followed by calculating the $wTO$ for a time series where the observations are not independent of each other.}

\subsection*{A method for determining a Consensus Network}
Berto and collaborators \cite{berto2016consensus} described a consensus network based on gene-expression data from primates' frontal lobes by applying a Wilcoxon test on the links.  
Our proposed  methodology allows {the use of two or more datasets, each generating different (and significant) $wTO$ values,  to be combined into a single $CN$. Our approach}  has the advantage of penalizing links with opposite signs. According to the same rationale, links with the same sign among the multiple $wTO$ networks, will have their $CN_{i,j}$ values closer to the largest {$\omega_{i,j}$} of a link among the $w$ networks. Our first step is to remove nodes that do not exist in all networks. Consequently, if a node is absent in at least one network, we are not able to compute a consensus of the links that belong to that node. {It is particularly important not to associate factors that were not measured in a particular condition}.
  
In order to obtain {a single integrated network derived from multiple independent $wTO$ networks}, we calculate a $CN$ using the following approach: 

If we have {$k=1,\ldots,n$} replicated networks {(note that $n$ means the index of the networks, not the exponent of $\alpha$ nor $\omega$)}, then we define the consensus network   $\text{wTO}_{\textnormal{CN}} = [\Omega_{i,j}]$ as
\begin{align}
    \Omega_{ij} = \sum_{{k} = 1}^n  \alpha_{ij}^{{k}} \omega_{ij} ^{ {k}},
\end{align}
 where
\begin{align}
\alpha_{ij}^{ {k}} = \frac{|\omega_{ij}^{ {k}}|}{\sum_{{k}=1}^{n} |\omega_{ij} ^{ {k}}|}.
\label{eq:consensus}
\end{align}
A threshold can be used to remove links with $\Omega_{i,j}$ values close to zero, thus should not be included in the consensus network. {To join networks that were generated with the proposed wTO method into the consensus network, the p~values are combined using the Fisher method.}

\section*{Results and discussion} 

{The representation of interactions between a set of nodes by the $wTO$ method ~\cite{ravasz2002hierarchical,zhang2005general, carlson2006gene}  takes into account the overall commonality of all the links a node has, instead of basing the analysis only on calculating raw correlations among the nodes. It thus provides a more comprehensive understanding of how two nodes are related. Therefore, it is expected that a $wTO$ network contains more robust information about the connections among nodes than what would result from simply taking direct correlations into account~\cite{zhang2005general,nowick2009differences}. 
The $wTO$ can be computed based on a similarity matrix, where the link weights are calculated using Pearson's product moment correlation coefficient or the Spearman Rank correlation. The first one measures the linear relationship between two genes. Note that, the Pearson's correlation coefficient is sensitive to extreme values, and therefore it can exaggerate or under-report the strength of a relationship. The Spearman Rank Correlation is recommended when data is monotonically correlated, skewed or ordinal, and it is less sensitive to extreme outliers than the Pearson coefficient~\cite{altman1990practical, mccrum2008correct, mukaka2012guide, bishara2012testing}. }

\subsection*{Package functions}

The function \texttt{wTO} calculates the weights for all links according to Eq.~(\ref{eq:wto}) between a set of nodes for a given input data set. If the user is not interested in the resampling option, one may simply run this $wTO$ function.

To test whether the calculated $wTO$ is different from random expectation and to decide on a suitable threshold value for including link weights, we implemented the function \texttt{wTO.Complete}.  Here, the $wTO$ is calculated a number of times, $n$ specified by the user, by using either the 1) Bootstrapping (\texttt{method\_resampling $=$ ``Bootstrap''}), or (\texttt{method\_resampling  $=$ ``BlockBootstrap''}) for time series data  or 2) Permuting the expression values for each individual (\texttt{method\_resampling  $=$ ``Reshuffle''}) \cite{nowick2009differences}. The user may specify the correlation {method} that this function should use, Pearson correlation is the default choice. 

Because bootstrapping and permutation tests can be computationally expensive, the \texttt{wTO.Complete} can also run in parallel over multiple cores to reduce the wall clock time.  For running in parallel, the user may specify a given number of $k$ computer threads to be used in the calculations. To implement the parallel function, we used the \texttt{R} package \texttt{parallel}~\cite{stats}.

The execution of the \texttt{wTO.Complete} function returns two outputs; a diagnosis set of plots and a list consisting of the following three objects:
\begin{itemize}
\item \texttt{\$Correlation} is a data.table containing the Pearson or Spearman correlations between all the nodes, not only the set of interest. The $wTO$ links for the set of nodes of interest are based on these correlations. The default of this output is set to \texttt{FALSE}.

\item \texttt{\$wTO} is a data.table containing the nodes, the $wTO$ values (signed and unsigned), the p-values and the adjusted p-values computed using both signed and unsigned correlations.

\item \texttt{\$Quantile} is a table containing the {quantiles for the empirical distribution, computed using the bootstrap and the quantiles for the real data}: 0.1\%,  2.5\%,   10\%,  90\%, 97.5\% and 99.9\%. Those empirical values can be used as a threshold for the $wTO$ values, when it is not desired to visualize low $wTO$ scores.
\end{itemize}
The set of plots indicate the quality of the resample: the closer the density of the resampled data is to the real data, the better. Another generated plot is the scatter plot of the ${\omega_{i,j}}$ vs $|{\omega_{i,j}}|$, as previously discussed. The scatter plot of p-values against the ${\omega_{i,j}}$ and $|{\omega_{i,j}}|$ is also plotted along with  suggested threshold values that are the empirical quantiles. 

\begin{figure}[ht]
	\centering
	\includegraphics[width=\textwidth]{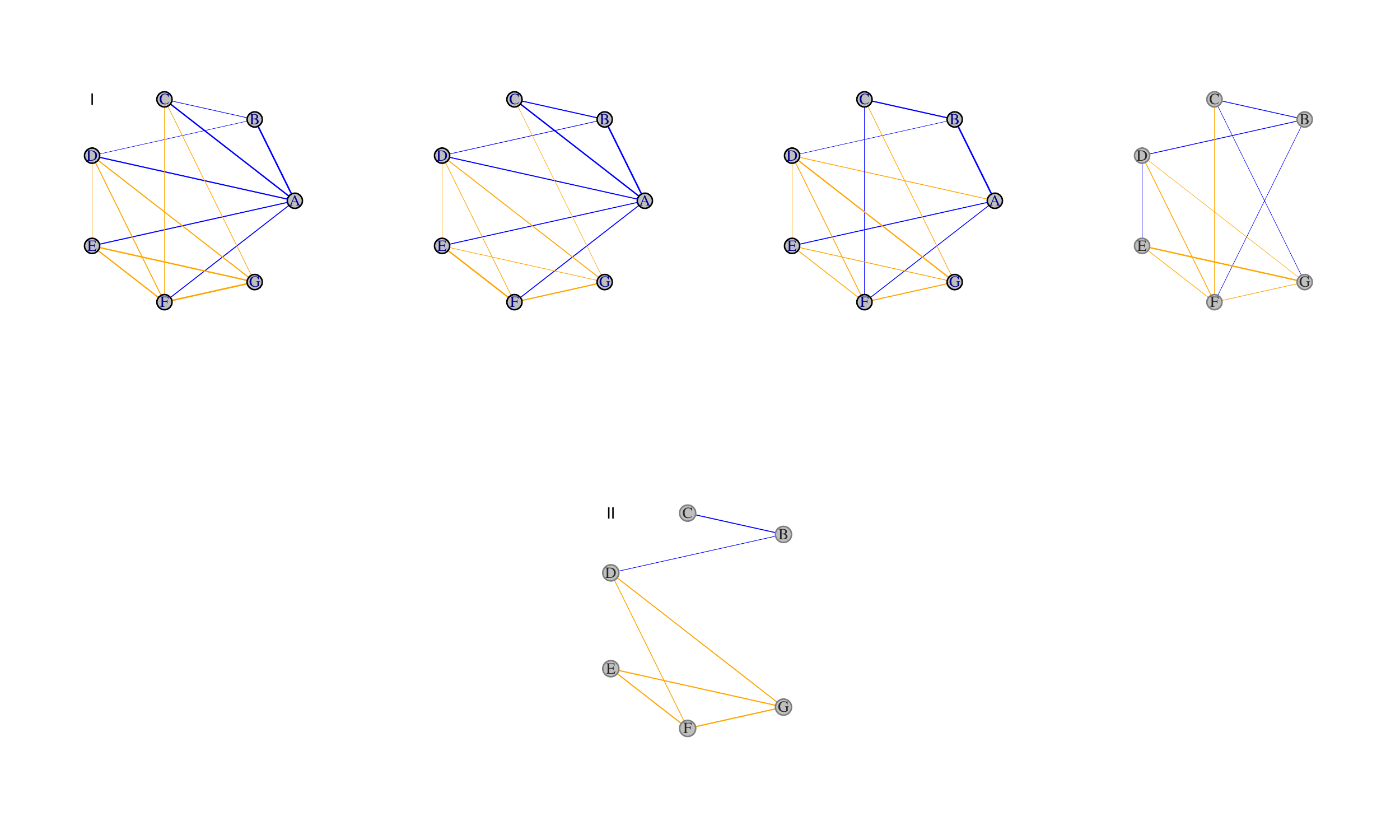}
	\caption{\textbf{A schematic example of the $CN$ method}: Panel I shows four independent networks to be combined into one $CN$. Note that the rightmost network does not include the 'A' node. Blue links indicates negative sign, while orange, positive. The $CN$ can be seen on Panel II. Note that the missing node from Panel I is not present in the $CN$. Also, only links that are constant in its sign among networks are present in the final network. For example, the link between D and E is removed since it has a different signal in the last network.}
	\label{figure:CNmethod}
\end{figure}

{Computing} of the $CN$ is done using the function \texttt{wTO.Consensus}. This function allows the user to give a list of networks in the format of data.frames with: Node 1, Node 2, the link weight {and the p-value}. The output is a data.table containing the two nodes' names and the consensus weight, {and the combined p-value. This allows the user to filter out the links that were not significant in part of the networks}.  
{A visual representation of the Consensus Network methodology is shown in Fig. ~\ref{figure:CNmethod}. The thicker the link between two nodes is, the stronger the correlation between them. The signs are represented by the colors blue and orange, respectively. If a link has different signs in the networks, the strength of the link in the $CN$ is close to zero. When all links agree to the same value or show little deviation, the strength of the resulting $CN$ value is closer to the determined $|maximum|$ value. If a node is absent in at least one network, it is removed.}

{The output data.frames (from both, \texttt{wTO.Complete} and \texttt{wTO.Consensus}) can be easily exported using the function \texttt{export.wTO}. This allows, for instance, to pass on the results of our package to Cytoscape \cite{shannon2003cytoscape} for further analysis.} 

Our \texttt{R} package also includes options to visualize the resulting networks. The function \texttt{NetVis} generates an interactive graph using as input a list of links and their corresponding weights. The analysis functions \texttt{wTO.Complete} and \texttt{Consensus} both generate network data-structures (edge list) that can be visualized with this function. 
The user needs to choose a relevant $wTO$-threshold {(the quantiles resulting from the bootstrap)}, or $p$-value cut-off, to select the set of links to be plotted. Additionally, the user may choose a layout for the network visualization from those available in the \texttt{igraph} \cite{csardi2006igraph} package.  By default, the $wTO$-threshold value is set to $0.5$, and the network layout-style is set to \texttt{layout\_nicely}. {To avoid false positives, we recommend to filter the data according to the desired significance $p$~value and to choose the $wTO$-threshold according to the computed empirical quantiles. }
The size of the nodes is relative to their degree. 
Our package further includes an option for making clusters from the nodes; {if allowed, nodes are colored according to the cluster they belong to. The user can choose the method to create the clusters.}

{One important difference between our package and the \texttt{WGCNA} package, is that we only use significant links for cluster (modules) network representation} instead of the full set of co-expressions, as in the \texttt{WGCNA}  package.
The width of a link is relative to the $wTO_{i,j}$, and its color is respective to its sign (if a signed network was calculated). Nodes can have different shapes, allowing for labeling nodes of different classes, for example target genes or protein coding and non-protein coding genes.
Furthermore, the user may also zoom in and out of the network visualization, drag nodes and links, edit nodes and links, and export the image as html or png. 
The package provides example datasets and an example of nodes of interest as well. 

\subsection*{Algorithm compute time with varying system size}

\begin{figure}[h!]
	\centering
	\includegraphics[width=\textwidth]{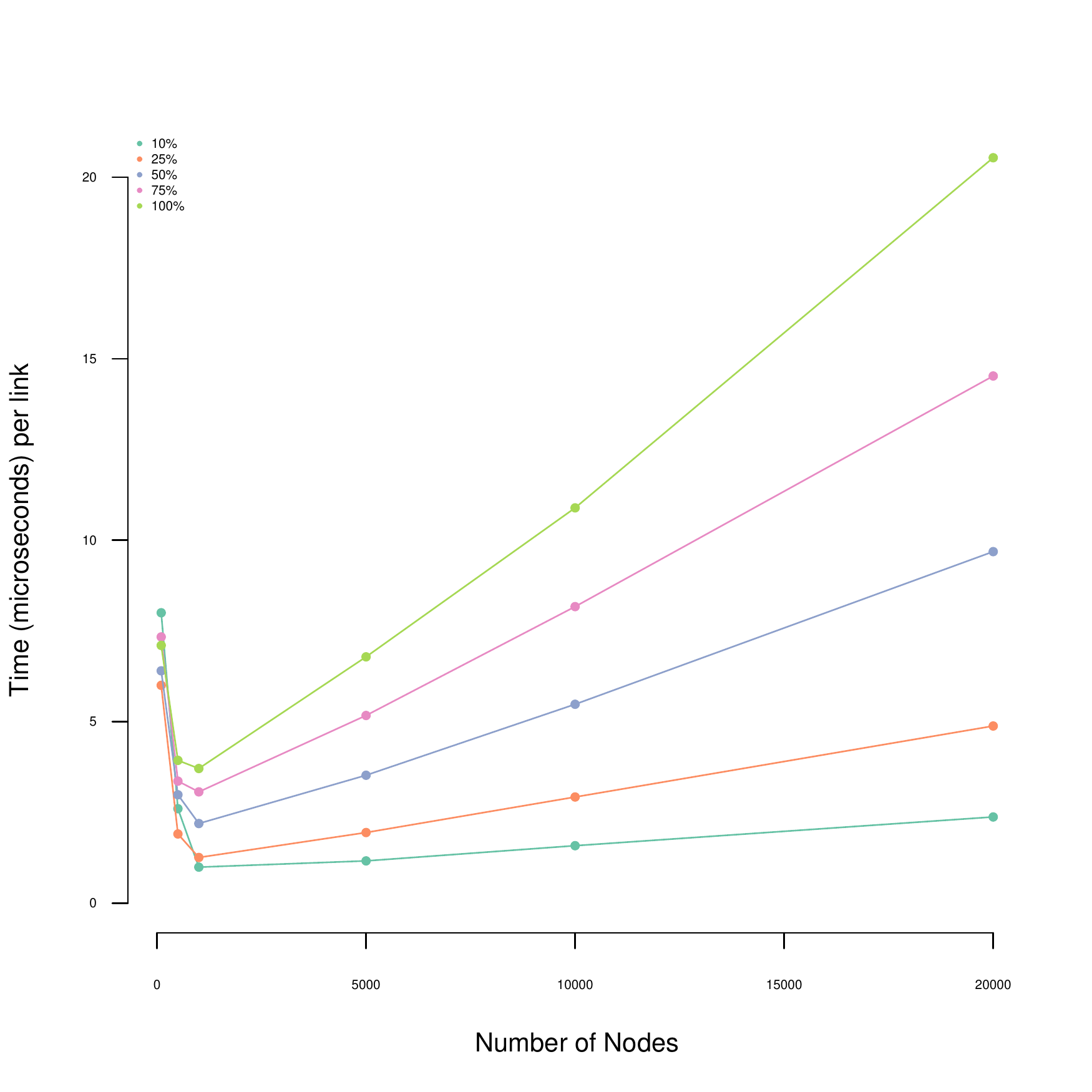}
	\caption{\textbf{Computational time for the calculation of \texttt{wTO} for each link for different sizes of networks and proportions of sets of nodes of interest:} The run time of the $wTO$ calculation increases with increasing proportion of nodes of interest. The graph presented here shows the time for computing each link for different sizes of nodes and proportions of subsets of nodes of interest.}
	
	\label{figure:Time}
\end{figure}

Normally, when running the $wTO$, the interest lies on a subset of nodes of interest. In Fig. \ref{figure:Time} we show the runtime for different network sizes, and different proportions of nodes of interest.
When running the $wTO$ for all expressed genes coding for {transcription factors (TFs)} being the genes of interest, we have around 14\% of nodes of interest. Using a standard laptop computer, it's possible to compute the $wTO$ for a full network with 20,000 nodes in 20 miliseconds per link. This shows that it is quite feasible to compute the full $wTO$ for a realistic gene expression network.  

\subsection*{Comparison with existing methods}

{A variety of methods currently exist to analyze gene co-expression networks, in particular \texttt{ARACNe} \cite{margolin2006reverse, margolin2006ARACNe}, SPACE~\cite{peng09} and \texttt{WGCNA} ~\cite{langfelder2008wgcna, fastR}. These methods rest on a multitude of different mathematical principles, particularly with respect to how co-expression is quantified. Of particular interest is \texttt{WGCNA}, which shares notable similarities with our \texttt{wTO}  package in {heuristic} terms, but with some substantial differences in functionality. 
In particular, \texttt{WGCNA}  also uses the weighted topological overlap (in their nomenclature, the ``topological overlap matrix", or TOM) to quantify co-expression at the gene-pair level. But in \texttt{WGCNA}, the final edge weight corresponds to the absolute value of $\omega_{i,j}$ as defined in Eq.~\ref{eq:wto}, or the absolute value of the terms in the numerator of Eq.~\ref{eq:wto}. These are referred to as signed or unsigned, respectively. Topological overlap as a measure of co-expression has previously been shown to compare favourably with other methods~\cite{Allen2012}. }

{While wTO and WGCNA construct the networks based on overlaping topologies, the ARACNe method builds the network using the mutual information (MI) and removing links that are indirect interactions using data processing inequality (DPI). Another important difference between the methods is that wTO and WGCNA will compute a link for all pair-wise possible connections, while ARACNe will only compute the pair-wise information if their information is not independent.}

{Relative to \texttt{WGCNA}, \texttt{wTO} provides three major additions: the determination of p-values (determined by bootstrapping) for each pairwise wTO value; the calculation of a consensus network, and the ability to visualize the topological overlap network (along with node grouping according to a choice of nine algorithms). 
While \texttt{WGCNA} provides a variety of tools for visualizing the hierarchical tree forming the network, as well as for rendering the correlation matrix in heatmap form, it does not provide a node-and-edge type view of the co-expression network (but does allow for exporting networks into Cytoscape, in which network views are possible). 
Additionally, the consensus network as defined in Eq.~\ref{eq:consensus} differs from the consensus TOM defined in \texttt{WGCNA}, which simply assigns to each edge of the consensus network the minimal value of the topological overlap across the input conditions. 
This is a strict version of consensus (unanimity), in that it will discard any gene pair if the overlap is weak in even a single network. 
In contrast, while Eq.~\ref{eq:consensus} will remove contributions from networks where the topological overlap is weak (or where the sign of the wTO score is in conflict with the other networks), an edge may still be included if it is sufficiently present across the other networks.}

{Further additions in \texttt{wTO} include the possibility of choosing the Spearman correlation as the basis of $a_{i,j}$ (while \texttt{WGCNA} provides biweight midcorrelation, or bicor for short; both provide Pearson), as well as reducing computation time by the option of restricting the calculation of wTO scores to a set of genes of interest (while still including the adjacency to genes outside this set in each inter-set wTO score). }

{Another minor difference resides in how wTO is determined for each gene with itself. From Eq.~\ref{eq:wto}, we see that (assuming $a_{i,{i}} = 0$ and $a_{i,j} = a_{j,i}$):}

\begin{align}
\omega_{i,i} = \frac{\sum_{u=1}^N a_{i,u}a_{u,i} + a_{i,i}}{k_i + 1 - |a_{i,i}|} = \frac{\sum_{u=1}^N a^2_{i,u}}{\sum_{u=1}^N a_{i,u} + 1}.
\label{eq:wtoSelf}
\end{align}

{For an unweighted network, where $a_{i,j} = 0$ or $a_{i,j} = 1$ for all $(i,j)$, this approximates to $\omega_{i_i} \approx 1$ for large $k_i$. However, this is not the case for  weighted networks. \texttt{WGCNA} differs from the wTO package in that $w_{i,i} = 1$ is explicitly set for all $i$, while our package retains the score as defined by Eq.~\ref{eq:wto}.}

\begin{table}[h]
\scriptsize
\centering
\caption{Comparison of key differences between wTO,  \texttt{WGCNA} and \texttt{ARACNe}}
\label{COMPARISON}
\begin{tabular}{l|ccc}
\toprule
\textbf{Method} & \textbf{wTO} & \textbf{WGCNA} & \textbf{ARACNe} \\
\hline
Topological overlap & Yes & Yes & No\\

Signed topological overlap & Optional & No & No\\

Consensus topological overlap & Weighted sum & Minimum weight (strict) & No\\

Pairwise p-values & Yes & No & Used to filter MI\\

Network view & Native & Exported to Cytoscape & Exported to Cytoscape \\

Soft thresholding & No & Optional (on by default) & No\\

Correlation choices & Spearman, Pearson & Bicor, Pearson & Spearman, Pearson, Kendall \\

Able to deal with time-series & Yes & No & No \\
\hline

\end{tabular}
\end{table}

\subsection*{Comparing wTO, \texttt{WGCNA} and \texttt{ARACNe} using an \textit{E. coli} Transcription Factor network}

{In order to  quantitatively compare the performance of wTO, \texttt{WGCNA} and \texttt{ARACNe}, we downloaded a gene expression dataset from \textit{E. coli} from \url{http://systemsbiology.ucsd.edu/InSilicoOrganisms/Ecoli/EcoliExpression2} \cite{Lewis2009,Fong2005, Fong2006,Covert2004}. 
The data consists of 213 Affymetrix microarray gene expression profiles, corresponding to multiple different strains under different growth conditions, and contains gene expression data for 7312 distinct probes. 
Gene expressions were calculated as the mean of probes  corresponding to the same gene. 
To assess the capability of the three tools in identifying true TF-TF interactions, we used the RegulonDB \cite{Gama-Castro2016} database, which contains experimental data from \textit{E. coli}, as a reference.
{We defined as True-Positive interactions those that are described in RegulonDB, and as True-Negatives all interactions that could not be experimentally validated in that dataset.
For comparison, we also calculated  networks using only the raw Pearson correlation.} We generated the network for \texttt{WGCNA} following the steps described by the authors in the Tutorial \cite{zhang2005general,Horvath17402}. 
We used the functions \texttt{pickSoftThreshold} and \texttt{pickHardThreshold} for defining the power of the soft-threshold and for choosing the hard-threshold, respectively. 
The power was defined as 4 and the hard-threshold was set to 0.3.}

\begin{figure}[h!]
	\centering
	\includegraphics[width=\textwidth]{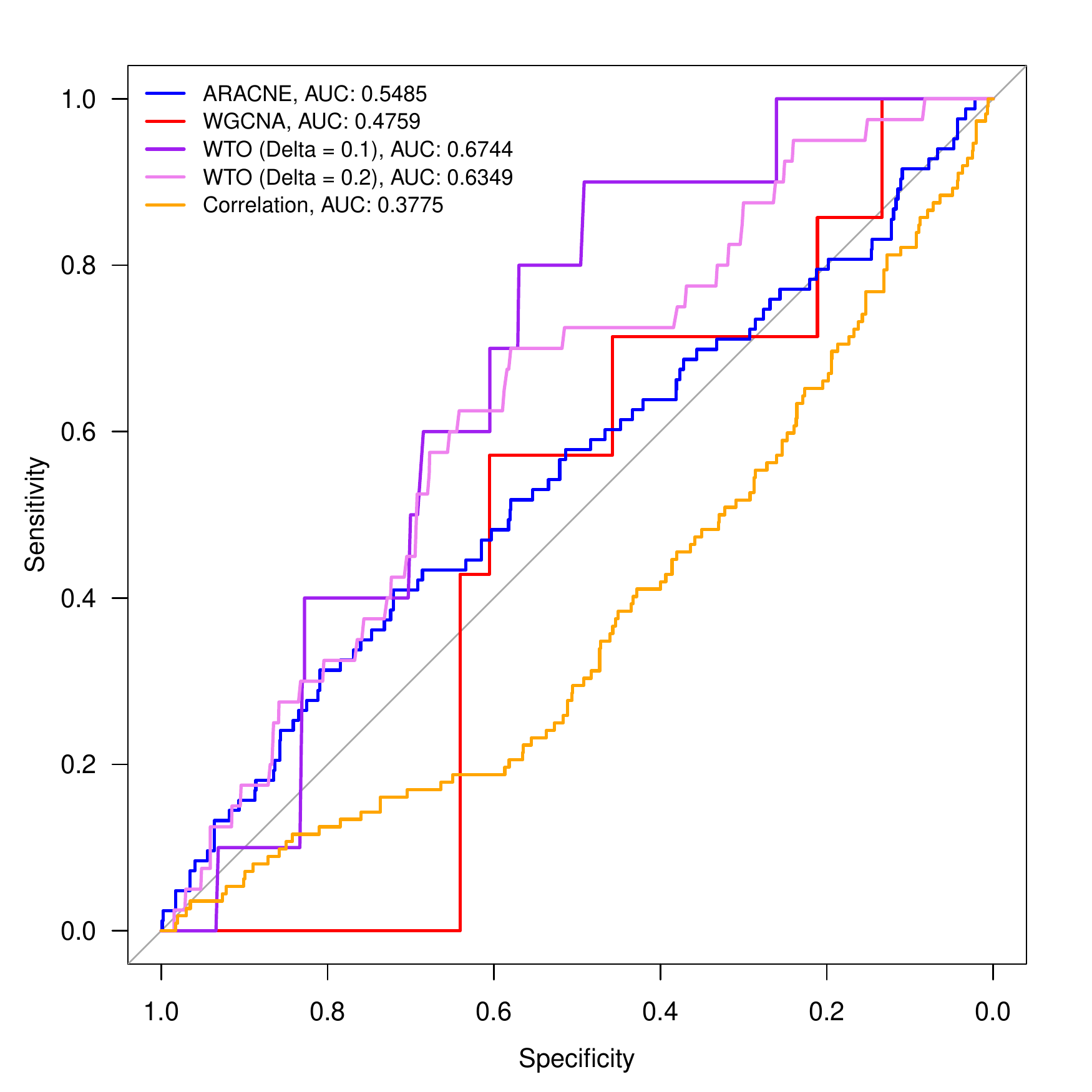}
	\caption{{\textbf{ROC curves for the comparison of methods.} Overall, our wTO method performs better than \texttt{ARACNe}, \texttt{WGCNA} and raw Pearson correlations. \texttt{ARACNe} is better in finding true positives, while \texttt{WGCNA} is more conservative, and therefore better in finding true negatives but identifies fewer true positives.}}
	\label{figure:ROC}
\end{figure}

{The \texttt{ARACNe} network was built using the Pearson correlation with \texttt{build.mim} and \texttt{ARACNe} functions in the \texttt{minet} R package \cite{ARACNeR}. The wTO network{s were} built using 100 simulations, Pearson correlation and filtered for p$_{adj}$~values $\leq $ 0.01 and the 90\% quantile. {One ARACNe network was constructed using a $\delta$ of 0.2, the default of the \texttt{package}, and another network was built using a $\delta$ of 0.1}. All networks were filtered to only contain TFs with information in the RegulonDB. We calculated the Receiver operating characteristic (ROC)-curve using the \texttt{pROC} R package \cite{pROC} (see Fig.~\ref{figure:ROC}).}

{ARACNE was able to better identify the amount of true positives compared to \texttt{WGCNA} and \texttt{wTO}, but performs worse when  finding true negatives and also has a larger number of false positives (Fig.\ref{figure:ROC}, Table \ref{Accuracy}). \texttt{WGCNA} is better at finding true negatives, but does not identify many true links. Our proposed wTO method performs better than \texttt{WGCNA} in finding true positives and better than \texttt{ARACNe} in finding true negatives. It also finds fewer false positives than \texttt{ARACNe}. In general, {even when using a large $\delta$,} wTO performs better than the two other methods, as seen in the Area Under the Curve (AUC; the closer it is to unity, the better).}
{This demonstrates that the use of the $wTO$ method further reduces false effects coming from incorrectly assigned linked genes (false positives) when compared to ARACNe and raw correlations.}

\begin{table}[h!]
\centering
\caption{Accuracy of the 3 methods and correlation}
\label{Accuracy}
\begin{tabular}{cllllll}\hline
\multicolumn{1}{l}{} & \multicolumn{1}{c}{\multirow{2}{*}{\textbf{\begin{tabular}[c]{@{}c@{}}ReactomeDB \\ (Total)\end{tabular}}}} & \multicolumn{1}{c}{\multirow{2}{*}{\textbf{\begin{tabular}[c]{@{}c@{}}Pearson \\ Correlation\end{tabular}}}} & \multicolumn{1}{c}{\multirow{2}{*}{\textbf{ARACNe}}} & \multicolumn{1}{c}{\multirow{2}{*}{\textbf{WGCNA}}} & \multicolumn{1}{c}{\multirow{2}{*}{\textbf{\begin{tabular}[c]{@{}c@{}}wTO\\ (delta 0.1)\end{tabular}}}} & \multicolumn{1}{c}{\multirow{2}{*}{\textbf{\begin{tabular}[c]{@{}c@{}}wTO \\ (delta 0.2)\end{tabular}}}} \\
\multicolumn{1}{l}{} & \multicolumn{1}{c}{} & \multicolumn{1}{c}{} & \multicolumn{1}{c}{} & \multicolumn{1}{c}{} & \multicolumn{1}{c}{} & \multicolumn{1}{c}{} \\ \hline
\multicolumn{1}{c|}{\textbf{True Negative}} & 7234 & 2259 & 2633 & 7092 & 6520 & 5235 \\
\multicolumn{1}{c|}{\textbf{False Negative}} & 0 & 216 & 245 & 321 & 318 & 288 \\
\multicolumn{1}{c|}{\textbf{False Positive}} & 0 & 4975 & 4601 & 142 & 714 & 1999 \\
\multicolumn{1}{c|}{\textbf{True Positive}} & 328 & 112 & 83 & 7 & 10 & 40 \\ \hline
\multicolumn{1}{l|}{\textbf{Total}} & 7562 & 7562 & 7562 & 7562 & 7562 & 7562
\end{tabular}
\end{table}
\section*{Examples of wTO networks using the wTO R package}
\subsection*{wTO and CN networks for TFs of the human prefrontal cortex}

{To exemplify the usage and results of our package, we analyzed three independent datasets of microarray data from human prefrontal cortex. Data sets were downloaded as raw data from Gene Expression Omnibus (GEO) website \cite{edgar2002gene}. From the study GSE20168 \cite{zhang2005transcriptional, zheng2010pgc},  we used data from a total of 15 postmortem brain samples.
From the study GSE2164 \cite{vawter2004gender}, we used a total of 26 samples from post mortem brains. 
And finally, from the study GSE54568 \cite{chang2014conserved} we used  all the 15 controls. All individuals were older than 5 years and died without any neuro-pathological phenotypes. We chose the TFs to be our genes of interest and calculated a TF-wTO network for each of the three datasets. Subsequently, we computed the consensus network for the three TF wTO networks.}

{The downloaded data were pre-processed and normalized by ourselves independently, using the \texttt{R} environment \cite{R}, and the \texttt{affy} \cite{affy} package from the \texttt{Bioconductor} set.
The probe expression levels (RMA expression values) and MAS5 detection $p$~values were computed, and only probesets significantly detected in at least one sample ($p$~value  $<$ 0.05) were considered. After the Quality Control and normalization of the data, the probes that were not specific for only one gene were deleted. If one gene was bound by more than one probeset, the average expression was computed.}

{Here, we will focus on how TFs are co-expressed in brain networks. We used a set of 3229 unique TF symbols from the TF-Catalog  (\textit{Perdomo-Sabogal et al. (in preparation)}) with ENSEMBL protein IDs. 
The construction of this catalog contains the information for TF proteins sourced from the most influential studies in the field of human GRF inventories \cite{messina2004orfeome, vaquerizas2009census, ravasi2010atlas, nowick2011gain, corsinotti2013global, tripathi2013gene, wingender2012tfclass, wingender2014tfclass} that are associated with gene ontology terms for regulation of transcription, DNA-depending transcription, RNA polymerase II transcription co-factor and co-repressor activity, chromatin binding, modification, remodeling, or silencing, among others.}

\begin{figure}[h!]
	\centering
	\includegraphics[width=\textwidth]{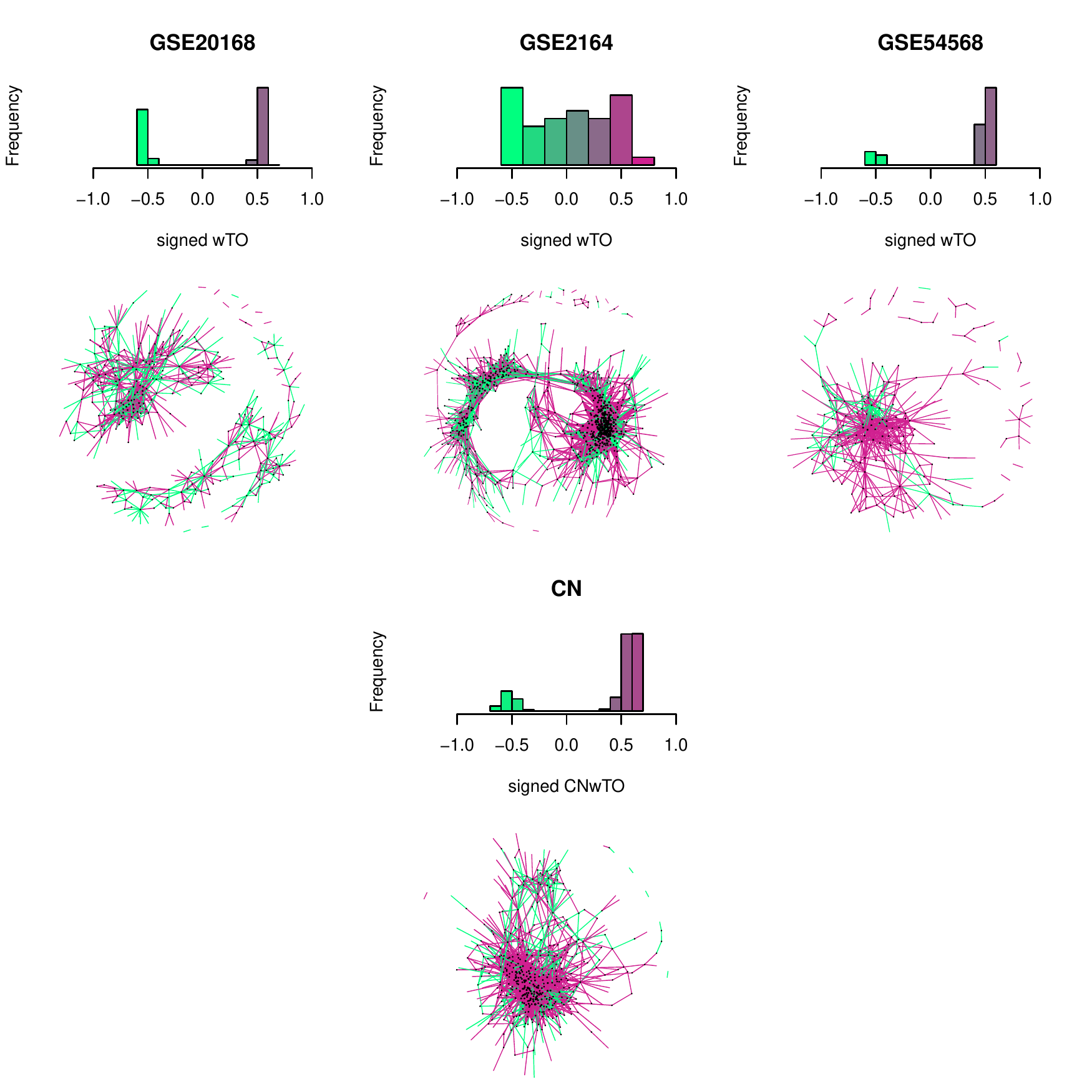}
	\caption{{\textbf{Comparison of the three networks used to compute the CN.} The first row shows the distribution of significant wTO values ($p_{adj}$~value $< 0.01 $). Note that the wTO range of the second network is larger than of the other two networks. The second row show the wTO network for each method. The third and forth row refer to the CN. Note that now the distribution of the wTO values does not include the wTO values close to zero, and retains only values that show a high correlation between the TFs. In the histograms, the presence of negative wTO values is visible, indicating that there are TFs that downregulate other genes.}}
	\label{figure:BrainNets}
\end{figure}

{
Signed wTO networks were calculated for each dataset separately using the function \texttt{wTO.Complete} of our \texttt{wTO} \texttt{R} package and then merged with the function \texttt{wTO.Consensus} into the consensus. Significance of all networks was evaluated using 1000 bootstraps, Pearson correlation and filtered for $p_{adj}$~value of $< 0.01$. The Consensus Network was built based on the calculated signed wTO values of significant links. Weights for links with in-significant wTO were set to zero. Fig. \ref{figure:BrainNets} shows the distributions and the networks for our three datasets.}

\begin{figure}[h!]
	\centering
	\includegraphics[width=\textwidth]{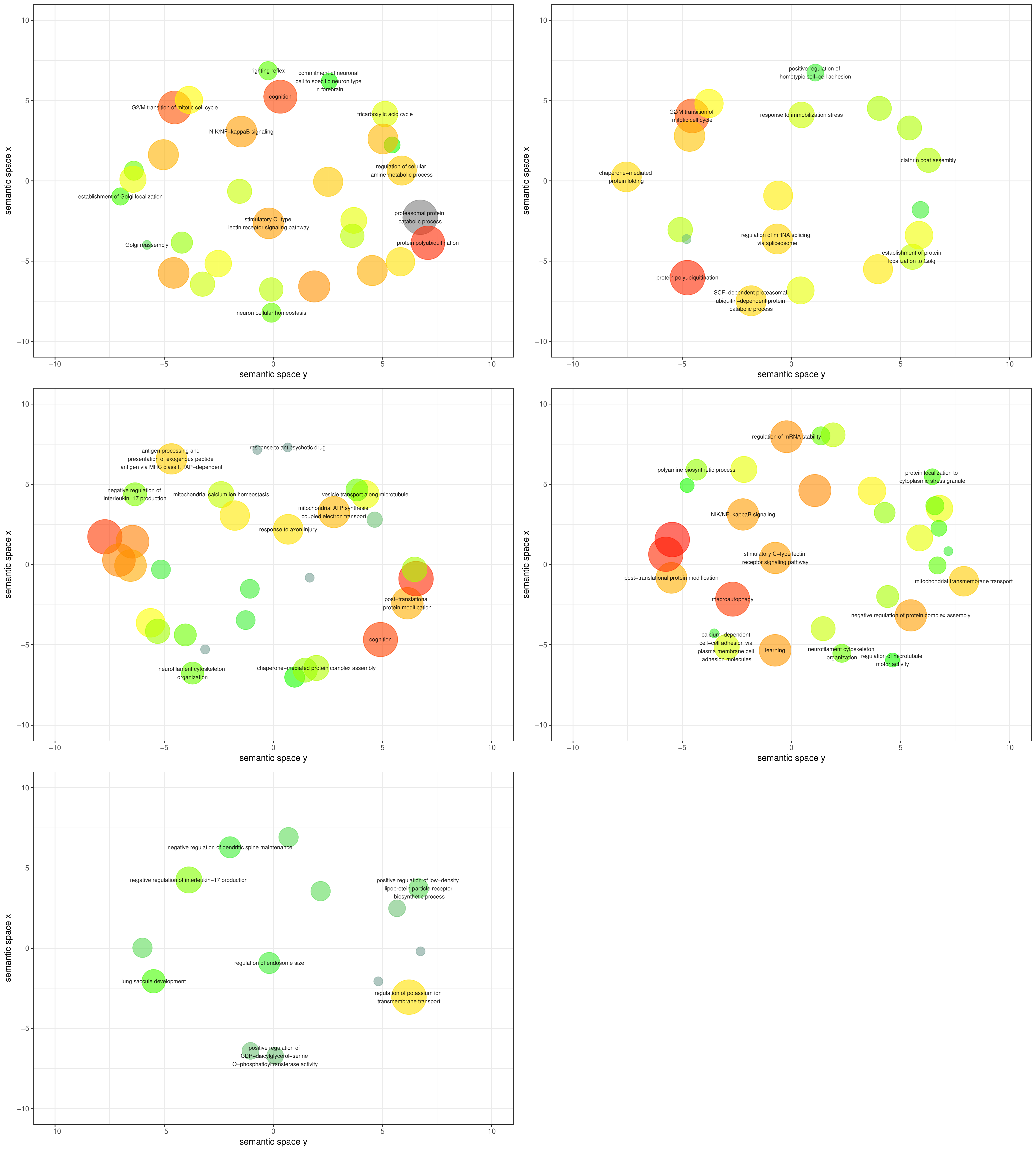}
	\caption{\textbf{GO terms enriched within each cluster.} Enriched GO terms of the category "biological process" are clustered by REVIGO [76] with the SimRel measurement and allowed similarity of 0.5. The size of the circle represents the frequency of the GO term in the database, i.e. GO groups with many members are represented by larger circles. The color code refers to the log$_{10}$(p-value) of the GO enrichment analysis: the closer to 0, the more red, the lower this value, the greener the bubble is. After removing redundancies, the remaining terms are visualized in \textit{semantic similarity-based} scatter-plots, where the axes correspond to semantic distance. Brain related functions were detected, for instance in Clusters 1 and 3, that are involved with cognition.}
	\label{figure:picGO}
\end{figure}

{TFs were clustered using the Louvain algorithm with the \texttt{NetVis} function, which identified 5 clusters in the CN. When considering each network independently, we had 18, 8 and 16 clusters.
This shows that the CN detects fewer clusters of genes, which are more densely connected{, compared to the clusters detected in the individual wTO networks.}
In order to investigate the function of each one of the 5 CN clusters, we calculated the correlation of each TF of a cluster with all other expressed genes using Pearson correlation. Genes with a correlation of at least $|0.80|$ with at least one TF of the cluster were used for GO enrichment analysis for that cluster, using the R package topGO \cite{topGO}. 
The enrichment analysis revealed many brain related functions, for instance, clusters 1 and 3 show overrepresentation of groups related to cognition  (Table \ref{GO} and Fig.\ref{figure:picGO}).
}
\begin{table}
\scriptsize
\centering
\caption{GO terms associated with each one of the CN Clusters}
\label{GO}
\begin{tabular}{@{}lllll@{}}
\toprule

\textbf{Cluster} & \textbf{TFs} & \textbf{Genes correlated to TFs} & \textbf{GO.ID} & \textbf{Term}                                   \\ \midrule
\multirow{31}{*}{1} & \multirow{31}{*}{589} & \multirow{31}{*}{58}                              & GO:0042775     & mitochondrial ATP synthesis coupled             \\
                 &               &                                  & GO:0010498     & proteasomal protein catabolic process           \\
                 &               &                                  & GO:0050890     & cognition                                       \\
                 &               &                                  & GO:0033238     & regulation of cellular amine metabolic pathway  \\
                 &               &                                  & GO:0008090     & retrograde axonal transport                     \\
                 &               &                                  & GO:0070050     & neuron cellular homeostasis                     \\
                 &               &                                  & GO:0090168     & Golgi reassembly                                \\
                 &               &                                  & GO:0006099     & tricarboxylic acid cycle                        \\
                 &               &                                  & GO:0051443     & positive regulation of ubiquitin-protein        \\
                 &               &                                  & GO:0061418     & regulation of transcription from RNA polimerase \\
                 &               &                                  & GO:0047496     & vesicle transport along microtubule             \\
                 &               &                                  & GO:0061640     & cytoskeleton-dependent cytokinesis              \\
                 &               &                                  & GO:0043488     & regulation of mRNA stability                    \\
                 &               &                                  & GO:0000086     & G2/M transition of mitotic cell cycle           \\
                 &               &                                  & GO:0038061     & NIK/NF-kappaB signaling                         \\
                 &               &                                  & GO:0000209     & protein polyubiquitination                      \\
                 &               &                                  & GO:0007052     & mitotic spindle organization                    \\
                 &               &                                  & GO:0031333     & negative regulation of protein complex          \\
                 &               &                                  & GO:0002223     & stimulatory C-type lectin receptor signal       \\
                 &               &                                  & GO:0016486     & peptide hormone processing                      \\
                 &               &                                  & GO:0034314     & Arp2/3 complex-mediated actin nucleation        \\
                 &               &                                  & GO:1900271     & regulation of long-term synaptic potential      \\
                 &               &                                  & GO:0000715     & nucleotide-excision repair, DNA damage          \\
                 &               &                                  & GO:1901983     & regulation of protein acetylation               \\
                 &               &                                  & GO:0016082     & synaptic vesicle priming                        \\
                 &               &                                  & GO:0043243     & positive regulation of protein complex          \\
                 &               &                                  & GO:2000637     & positive regulation of gene silencing           \\
                 &               &                                  & GO:0021902     & commitment of neuronal cell                     \\
                 &               &                                  & GO:0051683     & establishment of Golgi localization             \\
                 &               &                                  & GO:0060013     & righting reflex                                 \\
                 &               &                                  & GO:0061732     & mitochondrial acetyl-CoA biosynthetic pr...     \\ \midrule
                 \multirow{20}{*}{2} & \multirow{20}{*}{647} & \multirow{20}{*}{77} 
                              & GO:0035773     & insulin secretion involved in cellular          \\
                 &               &                                  & GO:0098930     & axonal transport                                \\
                 &               &                                  & GO:0000086     & G2/M transition of mitotic cell cycle           \\
                 &               &                                  & GO:0061640     & cytoskeleton-dependent cytokinesis              \\
                 &               &                                  & GO:0090083     & regulation of inclusion body assembly           \\
                 &               &                                  & GO:0034112     & positive regulation of homotypic                \\
                 &               &                                  & GO:1902750     & negative regulation of cell cycle G2/M          \\
                 &               &                                  & GO:0031146     & SCF-dependent proteasomal ubiquitin-dependent   \\
                 &               &                                  & GO:0061003     & positive regulation of dendritic spine          \\
                 &               &                                  & GO:0032922     & circadian regulation of gene expression         \\
                 &               &                                  & GO:0072600     & establishment of protein localization           \\
                 &               &                                  & GO:0061077     & chaperone-mediated protein folding              \\
                 &               &                                  & GO:0016191     & synaptic vesicle uncoating                      \\
                 &               &                                  & GO:1902309     & negative regulation of peptidyl-serine          \\
                 &               &                                  & GO:0048024     & regulation of mRNA splicing, via spliceosome    \\
                 &               &                                  & GO:0016486     & peptide hormone processing                      \\
                 &               &                                  & GO:0048268     & clathrin coat assembly                          \\
                 &               &                                  & GO:0000209     & protein polyubiquitination                      \\
                 &               &                                  & GO:0035902     & response to immobilization stress               \\
                 &               &                                  & GO:2000757     & negative regulation of peptidyl-lysine          \\ \midrule
                                  \multirow{31}{*}{3} & \multirow{31}{*}{402} & \multirow{31}{*}{17} 
                              & GO:0043687     & post-translational protein modification         \\
                 &               &                                  & GO:0050851     & antigen receptor-mediated signaling pathway     \\
                 &               &                                  & GO:0002479     & antigen processing and presentation             \\
                 &               &                                  & GO:0090199     & regulation of release of cytochrome c           \\
                 &               &                                  & GO:1905323     & telomerase holoenzyme complex assembly          \\
                 &               &                                  & GO:0050890     & cognition                                       \\
                 &               &                                  & GO:0043248     & proteasome assembly                             \\
                 &               &                                  & GO:0030177     & positive regulation of Wnt signaling pat...     \\
                 &               &                                  & GO:0047496     & vesicle transport along microtubule             \\
                 &               &                                  & GO:0042775     & mitochondrial ATP synthesis                     \\
                 &               &                                  & GO:0035773     & insulin secretion involved in cellular          \\
                 &               &                                  & GO:0045116     & protein neddylation                             \\
                 &               &                                  & GO:0090141     & positive regulation of mitochondrial            \\
                 &               &                                  & GO:0060071     & Wnt signaling pathway, planar cell              \\
                 &               &                                  & GO:0010635     & regulation of mitochondrial fusion              \\
                 &               &                                  & GO:0016579     & protein deubiquitination                        \\
                 &               &                                  & GO:0090090     & negative regulation of canonical Wnt signal     \\
                 &               &                                  & GO:0051131     & chaperone-mediated protein complex              \\
                 &               &                                  & GO:0051560     & mitochondrial calcium ion homeostasis           \\
                 &               &                                  & GO:0008090     & retrograde axonal transport                     \\
                 &               &                                  & GO:0032700     & negative regulation of interleukin-17           \\
                 &               &                                  & GO:0048170     & positive regulation of long-term neuronal       \\
                 &               &                                  & GO:0051036     & regulation of endosome size                     \\
                 &               &                                  & GO:0061588     & calcium activated phospholipid                  \\
                 &               &                                  & GO:0090149     & mitochondrial membrane fission                  \\
                 &               &                                  & GO:0097112     & gamma-aminobutyric acid receptor                \\
                 &               &                                  & GO:0097332     & response to antipsychotic drug                  \\
                 &               &                                  & GO:0097338     & response to clozapine                           \\
                 &               &                                  & GO:1902683     & regulation of receptor localization             \\
                 &               &                                  & GO:0060052     & neurofilament cytoskeleton organization         \\
                 &               &                                  & GO:0048678     & response to axon injury                         \\ \midrule
                 \end{tabular}
\end{table}

\begin{table}
\scriptsize
\centering
\caption{Continuation: GO terms associated with each one of the CN Clusters}
\label{GO2}
\begin{tabular}{@{}lllll@{}}
\toprule
\textbf{Cluster} & \textbf{TFs} & \textbf{Genes correlated to TFs} & \textbf{GO.ID} & \textbf{Term}                                   \\ \midrule

                 \multirow{31}{*}{4} & \multirow{31}{*}{677} & \multirow{31}{*}{39}
                            & GO:0007612     & learning                                        \\
                 &               &                                  & GO:0000209     & protein polyubiquitination                      \\
                 &               &                                  & GO:0070646     & protein modification by small protein           \\
                 &               &                                  & GO:0035567     & non-canonical Wnt signaling pathway             \\
                 &               &                                  & GO:0038061     & NIK/NF-kappaB signaling                         \\
                 &               &                                  & GO:0090313     & regulation of protein targeting to membrane     \\
                 &               &                                  & GO:0016339     & calcium-dependent cell-cell adhesion            \\
                 &               &                                  & GO:0002223     & stimulatory C-type lectin receptor signal       \\
                 &               &                                  & GO:0043687     & post-translational protein modification         \\
                 &               &                                  & GO:0008090     & retrograde axonal transport                     \\
                 &               &                                  & GO:0061732     & mitochondrial acetyl-CoA biosynthetic           \\
                 &               &                                  & GO:0070050     & neuron cellular homeostasis                     \\
                 &               &                                  & GO:0016236     & macroautophagy                                  \\
                 &               &                                  & GO:0043488     & regulation of mRNA stability                    \\
                 &               &                                  & GO:0061178     & regulation of insulin secretion involved...     \\
                 &               &                                  & GO:0016486     & peptide hormone processing                      \\
                 &               &                                  & GO:0035493     & SNARE complex assembly                          \\
                 &               &                                  & GO:0034112     & positive regulation of homotypic                \\
                 &               &                                  & GO:1902260     & negative regulation of delayed rectifier...     \\
                 &               &                                  & GO:1902267     & regulation of polyamine transmembrane           \\
                 &               &                                  & GO:2000574     & regulation of microtubule motor activity        \\
                 &               &                                  & GO:0016082     & synaptic vesicle priming                        \\
                 &               &                                  & GO:0051560     & mitochondrial calcium ion homeostasis           \\
                 &               &                                  & GO:0006596     & polyamine biosynthetic process                  \\
                 &               &                                  & GO:0060052     & neurofilament cytoskeleton organization         \\
                 &               &                                  & GO:1903608     & protein localization to cytoplasmic stress      \\
                 &               &                                  & GO:0000715     & nucleotide-excision repair, DNA damage          \\
                 &               &                                  & GO:0047496     & vesicle transport along microtubule             \\
                 &               &                                  & GO:1990542     & mitochondrial transmembrane transport           \\
                 &               &                                  & GO:0031333     & negative regulation of protein complex          \\
                 &               &                                  & GO:0046826     & negative regulation of protein export           \\ \midrule
                 \multirow{14}{*}{5} & \multirow{14}{*}{18} & \multirow{14}{*}{4}
                              & GO:0072369     & regulation of lipid transport                   \\
                 &               &                                  & GO:1901379     & regulation of potassium ion transmembrane       \\
                 &               &                                  & GO:0032700     & negative regulation of interleukin-17           \\
                 &               &                                  & GO:0051036     & regulation of endosome size                     \\
                 &               &                                  & GO:1904219     & positive regulation of CDP-diacylglycerol       \\
                 &               &                                  & GO:1904222     & positive regulation of serine C-palmitoyl       \\
                 &               &                                  & GO:1905664     & regulation of calcium ion import                \\
                 &               &                                  & GO:2000286     & receptor internalization                        \\
                 &               &                                  & GO:0021769     & orbitofrontal cortex development                \\
                 &               &                                  & GO:0045716     & positive regulation of low-density lipo.        \\
                 &               &                                  & GO:0060430     & lung saccule development                        \\
                 &               &                                  & GO:0070885     & negative regulation of calcineurin-NFAT         \\
                 &               &                                  & GO:1900272     & negative regulation of long-term synaptic       \\
                 &               &                                  & GO:1902951     & negative regulation of dendritic spine          \\ \bottomrule
\end{tabular}
\end{table}

\subsection*{Time series: Metagenomics data from the ocean }

{Only about 1\% of marine bacteria can be easily studied using standard laboratory procedures \cite{mac2017effects}. This is a major drawback for the understanding of how those microorganisms interact. Systems biology methods can provide helpful insights  to shed  light on species interactions.}

{To demonstrate an application of our wTO package for time series data with no replicates, we use as an example  metagenomics data from The USC Microbial Observatory. The data is public available at \url{https://www.ebi.ac.uk/metagenomics/projects/ERP013549}.}

{The sampling site is located  between Los Angeles and the USC Wrigley Marine Laboratory on Santa Catalina and spans approximately 900 m of water. Over the course of 98 months, samples were taken once a month. Operational Taxonomic Unity (OTUs) were determined using 16S ribosomal RNA (rRNA). The authors found 67 OTUs that will be used in our analysis. In order to find the correct lag for the blocked bootstrap, we used the autocorrelation function (acf) for all OTUs and chose a median lag of 2. This allowed us to define the blocks with high autocorrelation in the same sample, meaning that for them the abundance of the OTU on each specific time point is correlated to the following next 2 time points.}

\begin{figure}[h!]
	\centering
	\includegraphics[width=\textwidth]{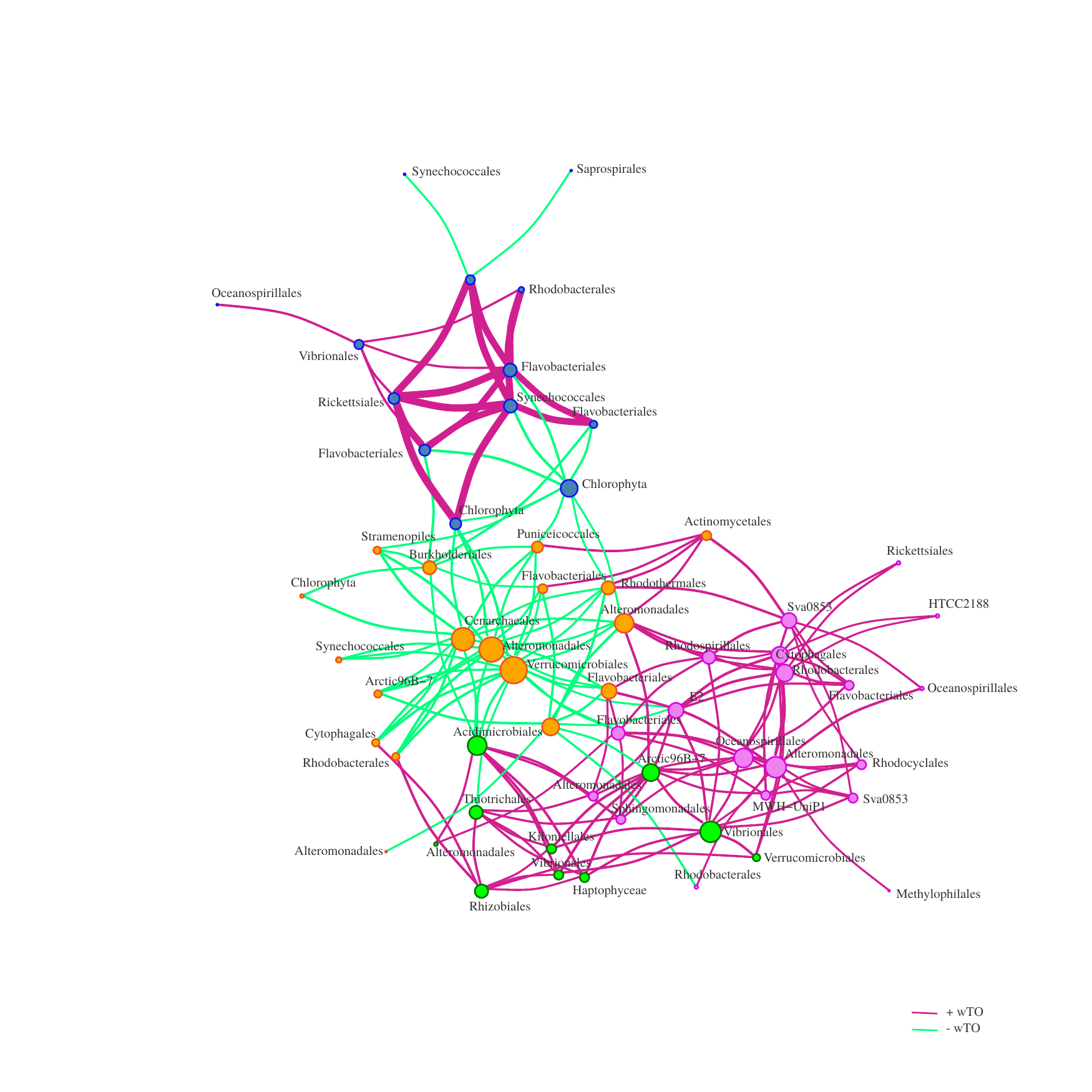}
	\caption{{\textbf{OTUs analysis using the Time-Series method of the \texttt{wTO} package}. In this network, the sizes of the nodes are proportional to a node's degree, and the width of a link is proportional to its wTO-absolute value. The link color refers to its sign, with green links being negative and purple ones positive. Nodes belonging to the same cluster are shown in the same color. There are four distinct clusters of bacteria. The orange cluster contains only negative interactions (green links), suggesting that the bacterial species in this cluster do not co-exist. We also notice, that many of the bacteria  belonging to the same order are well connected by purple links, indicating that they co-exist and share interactions. However, the number of interactions among non-related bacteria demonstrate that interactions are not intra-order specific.}}
	\label{figure:Metagenomics}
\end{figure}

{Based on that, we built the network of bacteria co-occurrence in that environment (Fig. \ref{figure:Metagenomics}). We found that 61 out of 67 OTUs had at least one significant interaction ($p_{adj}$~value $< 0.01$).} { Positive correlations in co-occurence networks may represent symbiotic or commensal relationships, while negative correlations may represent predator-prey interactions, allelopathy or competition for limited resources. Using the community detection method for defining clusters we identified four distinct clusters of bacteria. We did not find any association of the phylogeny with clusters, which is in agreement with previous studies. However, we can clearly see (Fig. \ref{figure:Metagenomics}) that the blue group is rich in negative relationships, while both, the purple and  orange groups, possess many positive relationships. These positive relationships are formed mostly by Flavobacteriales, bacteria that are known to infect fishes \cite{LOCH2015283} and to live in commensality with other bacteria from the same order \cite{bernardet2011flavobacteriales}.
}

\section*{Conclusion}
This new \texttt{wTO} package allows $wTO$ network calculation for both, positive and negative correlations, which is not provided in any other published \texttt{R} package. With this feature it becomes valuable for the analysis of gene regulatory network, { metabolic networks, ecological networks and other networks, in which the biological interpretation strongly depends on distinguishing between activating and inhibiting/repressing interactions}. 

Another novel feature is the computation of p-values for each link based on its empirical distribution, which allows for the reduction of false positive links {in wTO networks. With our package,} networks can also be calculated from time series data.
In addition, our package includes the computation of a $CN$, which enables integrating networks derived from different studies or datasets to determine links that consistently appear in these networks. 

By focusing on what these independently derived networks have in common, the CN should be of higher biological confidence than each individual network is. We also provide an interactive visualization tool that can be used to visualize both, $wTO$ networks and $CN$, for efficient further custom analysis.

{We qualitatively and quantitatively compared our new package to state-of-the-art methods and demonstrated that it performs better in identifying true positives and false negatives.}

{We provide two use cases for our package, one on wTO and CN calculation from three independent genome-wide expression datasets of human pre-frontal cortex samples, and one on wTO co-occurence networks calculated from time series data of a metagenomics abundance dataset from the ocean. Here, we demonstrated that clusters and GO enrichment in the CN are more defined than in individual wTO networks, highlighting the benefits of our package for analyzing and interpreting large biological datasets.}

\section*{Availability and requirements} \texttt{wTO} relies on the following packages: 	\texttt{som} \cite{yan2010package},  \texttt{plyr} \cite{plyr}, \texttt{stringr} \cite{wickham2010stringr}, \texttt{network} \cite{netart,netman},  \texttt{igraph} \cite{csardi2006igraph}, \texttt{visNetwork} \cite{visNetwork}, \texttt{data.table} \cite{data.table} and the standard packages \texttt{stats} and \texttt{parallel} \cite{stats}.
{The visualization tool implemented in our package was built  using a combination of the packages \texttt{network}~\cite{netart,netman}, \texttt{igraph} \cite{csardi2006igraph} and \texttt{visNetwork} \cite{visNetwork}}. 
{The \texttt{MakeGroups} parameter, passed to the function \texttt{NetVis} for constructing the network, allows the user to choose clustering algorithms from:  ``walktrap'' \cite{pons2006computing},   ``optimal''\cite{brandes2008modularity},   ``spinglass'' \cite{reichardt2006statistical, newman2004finding, traag2009community},   ``edge.betweenness' \cite{freeman1978centrality,brandes2001faster},   `fast\_greedy'' \cite{clauset2004finding}, ``infomap' \cite{rosvall2007maps, rosvall2009map}, ``louvain'' \cite{blondel2008fast}, ``label\_prop' \cite{raghavan2007near}   and   ``leading\_eigen'' \cite{newman2006finding}. All those algorithms are implemented in the \texttt{igraph} package\cite{csardi2006igraph}. 
}
  It is publicly available on CRAN repositories under the GPL-2 Open Source License \url{https://cran.r-project.org/web/packages/wTO/}. It is platform independent.

\section*{Declarations}

\subsection*{Abbreviations}
\textbf{wTO}: Weighted topological overlap;
\textbf{CN}: Consensus Network;
\textbf{TF}: Transcription Factor;
\textbf{ncRNA}: non coding RNA;
\textbf{miRNA}: micro RNA.
{
\textbf{OTU}: Operational Taxonomic Unit.
\textbf{acf}: autocorrelation function.
\textbf{GEO}: Gene Expression Omnibus.
\textbf{PFC}: Pre-frontal Cortex.
\textbf{AUC}: Area under the curve.
\textbf{ROC}: Receiver operating characteristic.
\textbf{TOM}: Topological Overlap Matrix.
\textbf{ARACNe}: An Algorithm for the Reconstruction of Gene Regulatory Networks.
\textbf{WGCNA}: Weighted Correlation Network Analysis.
\textbf{MI}: Mutual Information.
\textbf{DPI}: data processing inequality.
}

\subsection*{Acknowledgements}
We thank Professor Martin Middendorf, Martina Hall and Marlis Reich for fruitful discussions on the methodology and suggestions on the package. 
We thank Alvaro Perdomo Sabogal for providing us the Transcription Factors list used to build the PFC networks.
We thank Daniel Gerighausen for discussions.
  
\subsection*{Author's contributions}
    DG implemented the code in R. DG and TM conceived the idea of p-values for the edges. KN and EA generalized the $wTO$ for signed values. { DG and AV compared the wTO method to other methods. DG run the example analysis.} DG wrote the draft of manuscript. All authors discussed the manuscript, read and approved the final version of the manuscript.

\subsection*{Availability of data and materials}
     
     \texttt{wTO} is open source and freely available from \texttt{CRAN} \url{https://cran.r-project.org/web/packages/wTO/} under the the GPL-2 Open Source License. It is platform independent.
     
\subsection*{Competing interests}
  The authors declare that they have no competing interests.
\subsection*{Consent for publication}
Not applicable.

\subsection*{Ethics approval and consent to participate}
Not applicable.
\subsection*{Funding}
This work was supported partially by a doctoral grant from the Brazilian government's Science without Borders program (GDE 204111/2014-5).

%\bibliographystyle{plainnat} % Style BST file (bmc-mathphys, vancouver, spbasic).
%\bibliography{bmc_article}      % Bibliography file (usually '*.bib' )

\end{document}